\documentclass[intlimits,twoside,a4paper]{article}
\usepackage[cp1251]{inputenc}
\usepackage[eqsecnum]{cmpj3}
\usepackage{graphicx}
\usepackage{enumitem}

\articletype{A part of the Special collection to the 100th anniversary of the birth of Ihor Yukhnovskii}

\issue{2025}{28}{2}{23301}
\doinumber{10.5488/CMP.28.23301}
\title[Elastic properties of fluid mercury]%
{Elastic properties of fluid mercury across the metal-nonmetal
transition: {\it Ab initio} simulation study}
\author[T. Bryk, O. Bakai, A. P. Seitsonen]{T. Bryk\orcid{0000-0002-4360-0634}\refaddr{label1,label2}\thanks{Corresponding author: \email{bryk@icmp.lviv.ua}.},
        O. Bakai\refaddr{label3},
        A. P. Seitsonen\orcid{0000-0003-4331-0650}\refaddr{label4}}
\addresses{
\addr{label1} Institute for Condensed Matter Physics of the National Academy of Sciences of Ukraine, 79011 Lviv, Ukraine
\addr{label2} Institute of Applied Mathematics and Fundamental Sciences, Lviv Polytechnic National University, 79013 Lviv, Ukraine
\addr{label3} Akhiezer Institute for Theoretical Physics, National Science Center 
              ``Kharkiv Institute of Physics and Technology'' of NAS of Ukraine, 61108 Kharkiv, Ukraine
\addr{label4} D\'{e}partement de Chimie, \'{E}cole Normale Sup\'{e}rieure,
24 rue Lhomond, 75005 Paris, France
}

\Keywords{elastic properties, liquids, sound propagation, ab initio molecular dynamics}

\date{Received April 09, 2025, in final form May 20, 2025}

\begin{document}

\maketitle

\begin{abstract}
We report an {\it ab initio} molecular dynamics study of fluid mercury at temperature 1750~K
 in the range of densities 7--13.5 g/cm$^3$. Along this isothermal line we performed an analysis of
total charge fluctuations, which make evidence of neutral 
atom-like screening in fluid Hg for densities less than 9.25 g/cm$^3$, which practically coincides 
with the
 emergence of the gap in electronic density of states. High-frequency 
shear modulus, high-frequency and adiabatic speeds of sound, shear viscosity, Maxwell 
relaxation time and dispersion of collective excitations are analyzed as a function of density 
along the isothermal line. 
\printkeywords
\end{abstract}

\section{Introduction}

Metal-nonmetal (MNM) transition in condensed matter \cite{Mot68} is one of the fascinating topics in modern 
physics. Numerous experimental, theoretical and simulation groups focus on understanding the 
features of MNM transition in solid and liquid states of matter. Especially interesting and 
less explored is the case of MNM transition in liquids 
 because of the complicated interpay of topologicallly disordered structure, 
slow and fast relaxation processes and specific electronic screening in the region of 
density/pressure corresponding to the MNM transition \cite{Hen,Bon04,Tam10,Wei96,Kaj20,Kob21}.
The increase of external pressure 
usually causes metallization of liquids like it was observed for fluid hydrogen 
\cite{Sca03,Nel06,Mor10}, 
although 
the proofs of the pressure-induced localization of electrons
in the interstitial region like it is in crystals exist in the liquid state too~\cite{Tam10,Rat07,Bry13a,Bry14}, 
and causes  
the drop of electric conductivity.
Another interesting case is the MNM transition in a metallic liquid caused by its expansion 
as it was reported in liquid Hg \cite{Kre97,Tam07,Inu07,Cal11,Cal11b}. The structure and dynamics 
of liquid Hg was extensively studied by experimental and simulation techniques since then~\cite{Mun98,Bov02,Hos02,Bov02b,Ish04,Bom06,Yam06,Hos07,Kob07,Bom09,Cal09,Ayr14}.

Among different metallic liquids, the case of MNM transition in expanding Hg is 
especially interesting. The van der Waals
two-parameter phenomenological equation is insufficient to describe the phase states
of Hg (as well as in other metallic fluids) in which the gas-liquid and metal-nonmetal
transitions interfere. The first attempt to consider the solution of this problem, 
arguing the coexistence of the 1-st order MNM transition and the gas-liquid transition, was made
by Landau and Zeldovich~\cite{Lan43}. Historically, their paper stimulated many experimental,
theoretical and simulation studies of the evolution of electronic and atomic structures and
properties of Hg (see, e.g., \cite{Rad87,Hab90,Hab93,Inu03,Mar09,Rul10}). 
For fluid mercury, several important features are revealed \cite{Bak18}: 
i) the MNM transition in Hg takes place but it has nothing to do with the 
1-st order phase transition; ii) the fluid possesses a 3-state heterophase structure consisting 
of  the mesoscopic liquid-like-metallic, liquid-like-semiconducting, and gas-like  
species (called fluctuons in \cite{Bak18});  
iii)~strong effect of the electron subsystem on the continuously transforming, depending on density, 
atomic patterns and properties of the fluid.
The 3-state mesoscopic multiparametric theory of the gas-liquid and MNM transitions in fluid
Hg was developed and applied to description of the properties of Hg in \cite{Bak18}, which 
resulted in the phase diagram of 
Hg (figure~18 in \cite{Bak18}) with the heterophase states. The conductivity and polarizability of
the heterophase states were presented as well.

Here we investigate the electronic properties and ionic dynamics in Hg apart from 
the gas-liquid critical point towards higher densities by means of the {\it ab initio} molecular 
dynamics (AIMD) simulations.
Many interesting features can be studied in the dynamics of liquid Hg. In \cite{Inu05} the inelastic 
X-ray scattering experiments on mercury in a wide range of densities revealed a
sharp increase of the positive sound dispersion (PSD; called ``fast sound'' in \cite{Inu05}) in the 
region of MNM transition. The PSD is a specific characteristic of
the viscoelastic transition in liquids, which reveals the transition from macroscopic 
hydrodynamic mechanism (due to conservation laws) of sound propagation to microscopic elastic
mechanism like in solids. The theory of PSD in fluids \cite{Bry10} revealed the role of stress and heat
fluctuations in PSD, although in the region of MNM transition, the origin of the observed
sharp increase of PSD can be much more sophisticated. So far there is no clear explanation
what caused the increase of PSD in fluid~Hg. 

Recently there appeared an {\it ab initio} study of pressure-induced transitions in fluid 
H~\cite{Bry20} with an analysis of charge fluctuations in the
transition region. It was found that the total charge-charge structure factor 
$S_{QQ}(k)$ \cite{Mas03}
revealed a nonmonotonous behavior of its long-wavelength asymptote that was ascribed to the 
emergence of unscreened ions in the transition region. It is obvious that in the critical
MNM transition region, the electron density should strongly fluctuate with the  corresponding effect
on the screening properties. The emerging unscreened ions with long-range Coulomb interaction can definitely
 make a strong effect on collective dynamics. It would be very useful to check whether the
tendency observed in charge fluctuations at pressure-induced transitions in fluid H~\cite{Bry20} will
be present in the MNM transition in fluid Hg.

The remaining part of the paper is organized as follows. In the next section we 
provide details of the AIMD simulations, section III reports the results on electronic density
of states and total charge structure factors, density dependences of high-frequency and
adiabatic speeds of sound, as well as dispersion relations and discussion on the PSD across the MNM transition in fluid~Hg. The last section summarises the findings of the present study.

\section{Methodology}

{\it Ab initio} simulations were performed by the VASP package~\cite{Kre93,Kre93_a,Kre96,Kre96b} 
using a system of 200 particles in cubic box subjected to periodic boundary conditions
in NVT ensemble. We performed simulations at different 9 densities of fluid mercury
at temperature $T=1750$~K:
from very dense system with $\rho=13.5$~g/cm$^3$ down to very expanded system with $\rho=7.0$~g/cm$^3$. 
The electron-ion 
interaction was represented by the projector augmented wave (PAW) method~\cite{Blo94,Kre99}, in which two $6s$ and ten
$5d$ electrons were treated as valence electrons. The need to explicitly treat $5d$ electrons and 
consequently about 1300 wave functions at each step of AIMD restricted our system to only 
200 particles. The exchange-correlation functional was chosen in the PBE form \cite{Per96} of generalized gradient approximation,
and only $\Gamma$-point  of the Brillouine zone was used in the description of the electron 
density and the electronic kinetic energy.
The time step was~8~fs.

At each thermodynamic point, we equilibrated the system over 5 ps in NVT ensemble followed by the 
production runs each of 20~000 configurations. For each configuration we saved atomic coordinates, 
velocities
and forces, as well as components of stress tensor and Fourier-components of electron density. 
The latter were needed to estimate the total charge structure factors in the region of MNM 
transition in expanded~Hg.

The analysis of elastic properties and estimation of dispersion relations for liquid Hg were
performed by means of macroscopic stress autocorrelation functions \cite{Han} and 
estimation of propagating
eigenmodes of the generalized Langevin equation \cite{Han,Boo} in the frames of the approach of 
generalized collective
 modes~(GCM)~\cite{deS88,Mry95}. The macroscopic adiabatic speed of sound, $c_{s}$, was estimated using a recently suggested 
methology \cite{Bry23} which requires the knowledge of the  static correlation function of diagonal 
components of stress tensor $\psi^L(t=0)$ and of the high-frequency speed of sound $c_{\infty}$:
\begin{equation} \label{cs}
 c_s=\sqrt{c_{\infty}^2-\psi^L(0)/\rho}~.
\end{equation}
Here, $\psi^L(0)=V\langle \bar{\sigma}_{zz}\bar{\sigma}_{zz}\rangle/k_{\rm B} T$ 
with $\bar{\sigma}_{zz}(t)=
\sigma_{zz}(t)-P$  being the fluctuating part of the diagonal component of stress tensor, 
 $P$ is the pressure, $V$ is the volume of the simulated system, and $k_{\rm B}$ is Boltzmann constant.
The high-frequency speed of sound $c_{\infty}$ was estimated from the
 wavenumber-dependent stress tensor correlations, obtained from the relation
\begin{equation}
{\dot J}^{L,T}(k,t)=-\ri k\sigma^{L,T}(k,t),
\label{dotJ}
\end{equation}
where $L$ and $T$ mean longitudinal and transverse components, overdot means the  first time derivative.
The time evolution of spatial Fourier-components of mass-current density was obtained in AIMD
simulations using the expression
\begin{equation}
{\bf J}(k,t)=\frac{m}{\sqrt{N}}\sum_{i=1}^{N}{\bf v}_i(t)\, \re^{-\ri{\bf kr}_i(t)},
\end{equation}
and their first time derivative were similarly estimated as 
\begin{equation}
{\dot {\bf J}}(k,t)=\frac{1}{\sqrt{N}}\sum_{i=1}^{N}
[{\bf F}_i(t)+\ri m({\bf kv}_i){\bf v}_i(t)]\re^{-\ri{\bf kr}_i(t)},
\end{equation}
where $m$ is the atomic mass of Hg, ${\bf r}_i(t)$, ${\bf v}_i(t)$ and ${\bf F}_i(t)$
are the particle trajectory, particle velocity and force acting on the $i$-th particle,
respectively. 
The long-wavelength asymptote of wavenumber-dependent normalized
second frequency moment of the $L$-current spectra function is directly related to the 
high-frequency speed of sound as
\begin{equation} \label{cinf}
c_{\infty}\stackrel{k\to 0}{=}\frac{1}{k}\left[\frac{\langle {\dot J}^L(-k)\dot{J}^L(k)\rangle}
{\langle J^L(-k)J^L(k)\rangle}\right]^{1/2}~.
\end{equation}
The dispersion of extended acoustic collective excitations $\omega_s(k)$ was obtained in two ways: 
i) numerically via peak positions of the current spectral function $C^L(k,\omega)$ which is 
a numerical Fourier transformation of the AIMD-derived time correlation functions
$F^L_{JJ}(k,t)=\langle J^L(-k,t)J^L(k,t=0)\rangle$; and ii)~theoretically via imaginary 
part of the complex eigenvalues $z_{\alpha}(k)=\sigma(k)\pm \ri\omega_s(k)$, where $\sigma(k)$
is the $k$-dependent damping. These eigenvalues were obtained from the $5\times 5$ 
generalized hydrodynamic matrix ${\bf T}(k)$ generated for each $k$-point using the set of 
$5$ dynamic variables of the thermo-viscoelastic (TVE) dynamic model \cite{Mry95,Bry10} 
\begin{equation} \label{a5}
{\bf A}^{\rm(TVE)}(k,t) = \left\{n(k,t), J^L(k,t), \varepsilon(k,t),
\dot{J}^L(k,t), \dot{\varepsilon}(k,t)\right\},
\end{equation}
where the spatial Fourier components of particle density are
\begin{equation} \label{dynhyd1}
n(k,t)=\frac{1}{\sqrt{N}}\sum_{i=1}^{N}\re^{-\ri{\bf kr}_i},
\end{equation}
and the matrix elements involving fluctuations of the energy density $\varepsilon(k,t)$ and of its 
first time derivative $\dot{\varepsilon}(k,t)$ were treated in connection with AIMD simulations
according to the methodology suggested in~\cite{Bry13,Bry23b}.

\section{Results and discussion}

\begin{figure}[htb]
	\centerline{\includegraphics[width=0.48\textwidth]{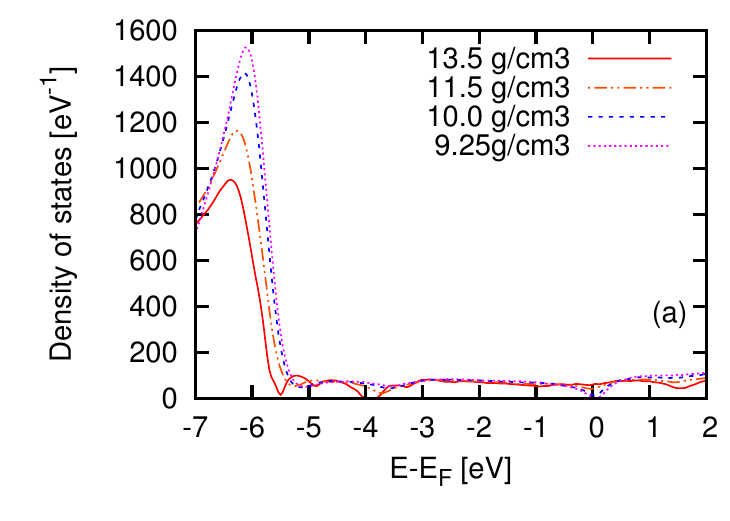}
	\includegraphics[width=0.48\textwidth]{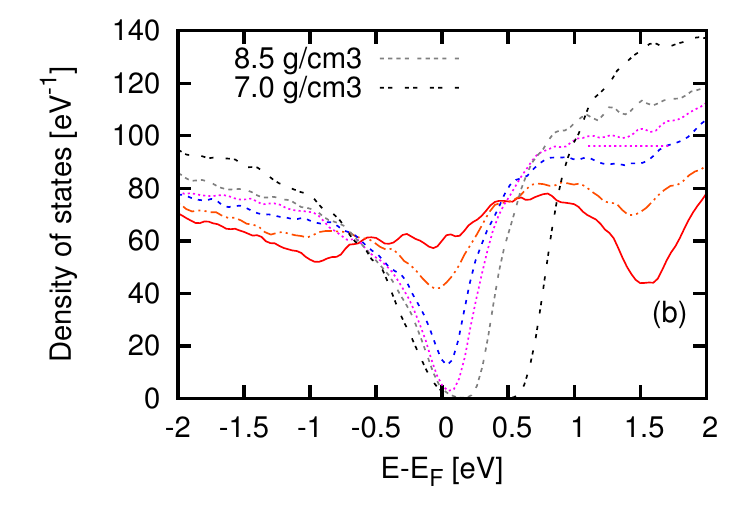}}
	\caption{(Colour online) Electron density of states for Hg fluid at several densities (a) 
		and in the vicinity of Fermi level (b).} \label{DOS}
\end{figure}

In order to locate the metal-nonmetal transition in fluid Hg we evaluated at each studied phase state the electronic density of states (EDOS) during the AIMD simulations. The energy
region $\sim $6--7~eV below the Fermi level is occupied by $5d$ electrons, while all the features 
of EDOS close to Fermi level are defined by $6s$ and $6p$ electrons. In figure~\ref{DOS} one can see that 
a tendency to form a pseudo-gap (a dip in EDOS at Fermi level) is observed already at density 
11.5~g/cm$^3$ while the gap emerges at density 9.25~g/cm$^3$ and increases with expanding  
the system. Since the $6s$ electrons constitute the EDOS close to the Fermi level,  the emergence of the gap 
might be a consequence of $6s$ electron localization in Hg neutral atoms.

\begin{figure}[h]
	\centerline{\includegraphics[width=0.48\textwidth]{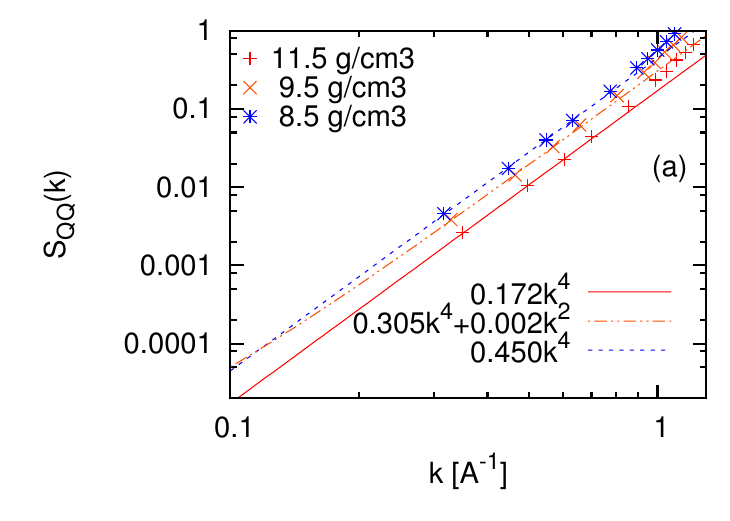}
		\includegraphics[width=0.48\textwidth]{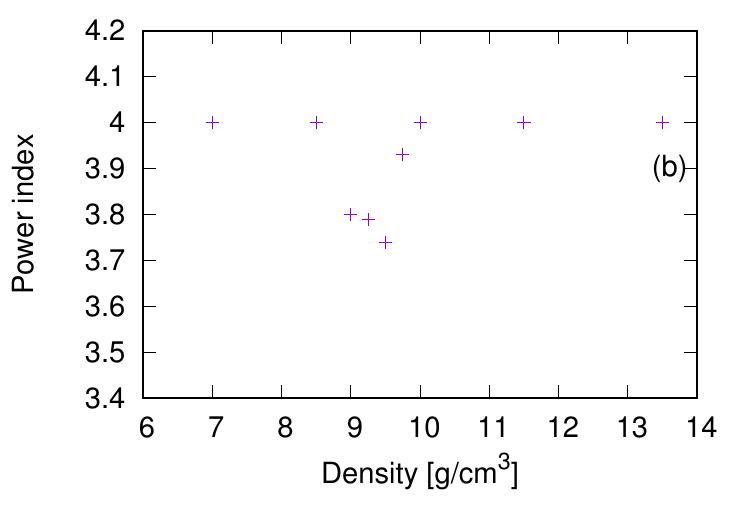}}
	\caption{(Colour online) Long-wavelength behavior of the charge-charge structure factor $S_{QQ}(k)$
		for three densities of liquid Hg at 1750 K (a) and the apparent exponent of the 
		long-wavelength asymptote of charge-charge structure factors
		at the smallest wave numbers for all densities studied  (b). }\label{index}
\end{figure}

An important information about the screening of ions by electron subsystem can be observed 
in the long-wavelength behavior of the total charge structure factors $S_{QQ}(k)$ \cite{Bry20}.
Here the Fourier components of
total charge density are represented as \cite{Mas03}

\begin{equation}\label{Qkt}
Q(k,t)=Z_{\rm ion}\,n_{\rm ion}(k,t)-n_{\rm el}(k,t)
=\frac{1}{\sqrt{N}}\bigg[Z_{\rm ion}\sum_{i}^{N}\re^{-\ri{\bf k}{\bf R}_i(t)}-n_{\rm el}(k,t)\bigg]~,
\end{equation}
where $N$ is the number of ions in the AIMD simulation cell,
and $n_{\rm ion/el}(k)$ are the Fourier components
of the ion/electron density corresponding to instantaneous ionic positions
$\{{\bf R}_i\}$. 
Recently it was found \cite{Bry20} that for the 
pressure-induced transformation from dielectric molecular fluid H$_2$ to metallic 
fluid hydrogen at high pressure, the long-wavelength asymptote of $S_{QQ}(k)$ reveals 
a non-monotonous dependence, which was suggested to be an evidence of  
under-screened ions existing in the transition region. The long-range Coulomb
interaction between under-screened ions should cause the $k^2$ long-wavelength asymptote 
of $S_{QQ}(k)$.
In figure~\ref{index} we show the observed long-wavelength asymptotes of the total 
charge factor $S_{QQ}(k)$ for Hg fluid at the studied densities along the isothermal line
$T=1750$~K. As one can see there is observed a behavior  very similar to the case of MNM transition
in fluid hydrogen. Right in the region of densities where the gap in electronic DOS
appears one observes a drop in the apparent exponent of $S_{QQ}(k\to 0)\sim k^{m}$, while
in the metallic region and at the two lowest densities one has a well-defined 
$\sim k^4$ asymptote.
Such a behavior of the exponent makes evidence of the existing under-screened ions 
in the region of the MNM transition in the fluid Hg.

\begin{figure}
	\centerline{\includegraphics[width=0.48\textwidth]{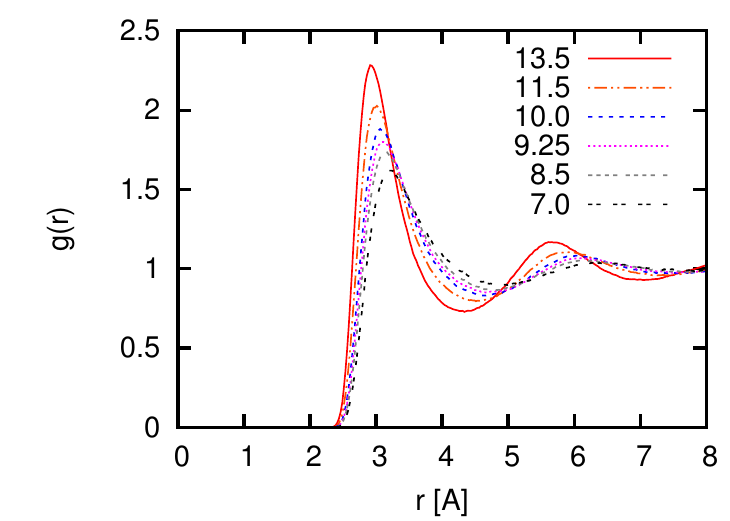}
		\includegraphics[width=0.48\textwidth]{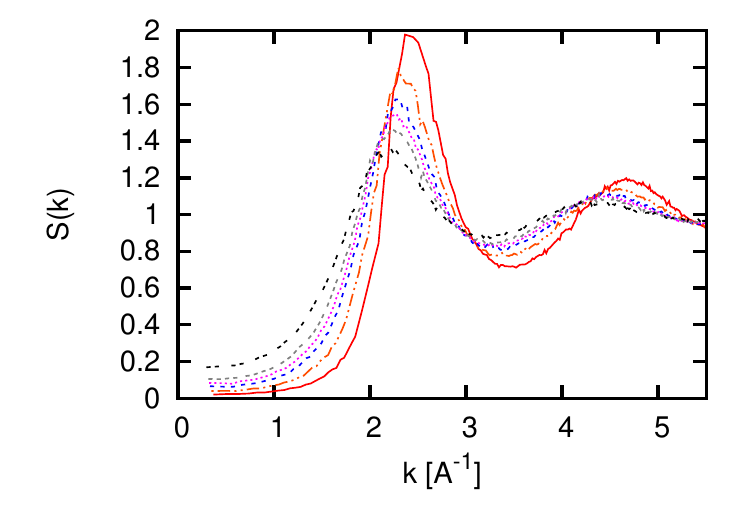}}
	\caption{(Colour online) Pair distribution functions $g(r)$ and static structure factors $S(k)$
		at six densities of liquid Hg at 1750~K. }\label{rdf}
\end{figure}

The structural properties such as pair distribution functions $g(r)$ and static structure factors
$S(k)$ are changing with density smoothly as one can see in figure~\ref{rdf}. No features were 
observed in the region of densities 9.0--9.5~g/cm$^3$, where the MNM transition is located, 
although the X-ray diffraction experiments~\cite{Tam98} indicated that possible emergence 
of Hg dimers can take place as the critical point on the phase diagram is approached.  In our 
AIMD simulations we did not observe more-less stable Hg associates at low densities.

We begin the analysis of the elastic properties with the 
shear-stress autocorrelation functions $\psi(t)$ for liquid Hg along the isothermal 
line $T=1750$~K, where $\psi(t)$ is defined as follows
$$
\psi(t)=\frac{V}{k_{\rm B}T}\langle\sigma_{xy}(t)\sigma_{xy}(0)\rangle. 
$$
The initial value of this function is the high-frequency shear modulus $G_{\infty}$
$$
G_{\infty}\equiv \frac{V}{k_{\rm B}T}\langle\sigma_{xy}(0)\sigma_{xy}(0)\rangle.
$$
Here, $\sigma_{ij}(t)$ are the fluctations of the stress tensor components during 
the MD runs,
and the shear-stress autocorrelation functions are defined via off-diagonal 
components of the
stress tensor, which fluctuate around zero value, i.e., $\langle\sigma_{xy}\rangle=0$.

\begin{figure}[h]
	\centerline{\includegraphics[width=0.48\textwidth]{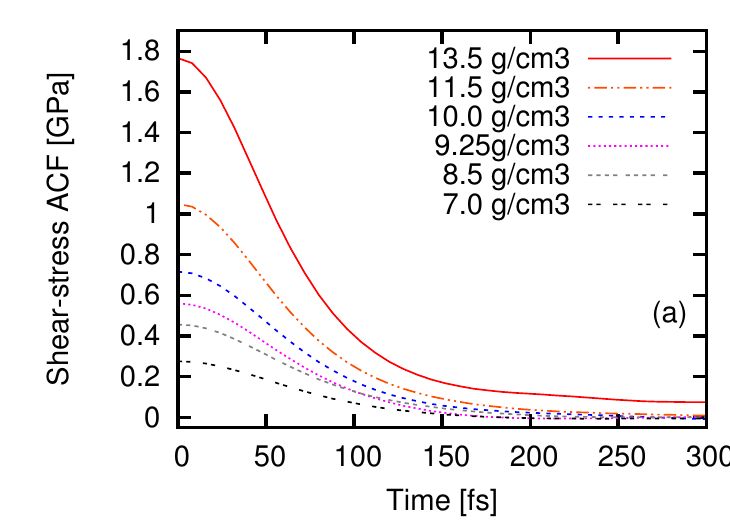}
		\includegraphics[width=0.48\textwidth]{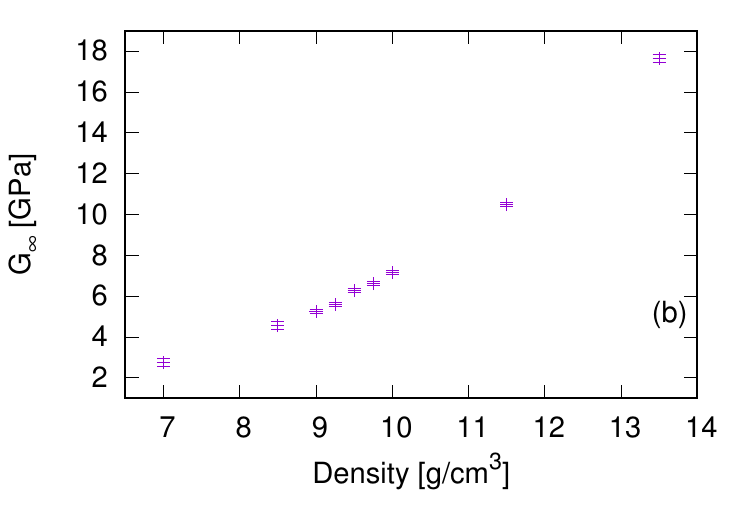}}
\caption{(Colour online) Density dependence of the shear-stress autocorrelation functions
$\psi(t)$ for liquid Hg at 1750 K (a) and of the high-frequency shear modulus
$G_{\infty}=\psi(t=0)$ (b). }\label{SACF}
\end{figure}

In the studied range of densities the $G_{\infty}$ 
increases monotonously 
with density (see figure~\ref{SACF}). On the other hand, the time dependence of the 
shear-stress autocorrelation
functions varies smoothly with density showing almost an 
exponential decay to zero. 
The only exception takes place for the density $\rho=9.25$~g/cm$^{3}$, for which the 
shear-stress autocorrelation function contains a negative tail while the short-time behavior  changes
smoothly with density and does not show any feature for that density point. 

\begin{figure}[h]
	\centerline{\includegraphics[width=0.48\textwidth]{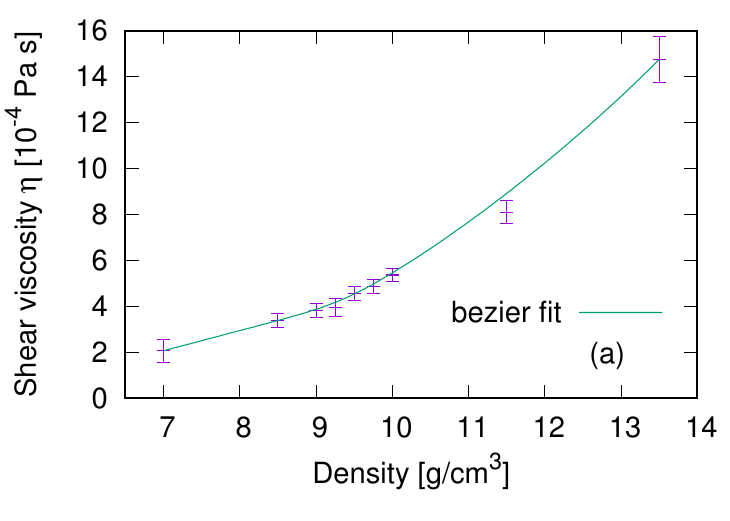}
		\includegraphics[width=0.48\textwidth]{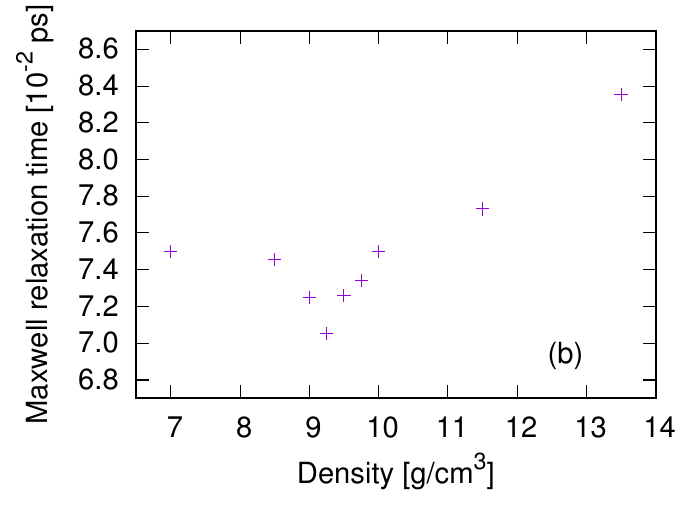}}
	\caption{(Colour online) Density dependence of the shear viscosity $\eta_s$ and Maxwell relaxation time $\tau_M$
		for liquid Hg at 1750 K. }\label{eta}
\end{figure}

The Green--Kubo integral over the shear-stress autocorrelation functions results 
in the shear viscosity~$\eta_s$
$$
\eta_s=\int_{0}^{\infty}\psi(t)\, \rd t
=\frac{V}{k_{\rm B}T}\int_{0}^{\infty}\langle\sigma_{xy}(t)\sigma_{xy}(0)\rangle\, \rd t.
$$
This integral implies that the shear viscosity at $\rho=9.25$~g/cm$^{3}$ 
should deviate from the smooth monotonous density dependence. Indeed, the calculated shear viscosities,
 shown in figure~\ref{eta}, show a slight drop of $\eta_s$ at $\rho=9.25$~g/cm$^{3}$. 

The Maxwell relaxation time in liquids, which is the characteristic time of 
shear stress relaxation, is defined by the ratio
$$
\tau_M=\frac{\eta_s}{G_{\infty}}\equiv\frac{1}{\psi(0)}\int_{0}^{\infty}\psi(t)\, \rd t\equiv \tau_{\rm corr}.
$$
In fact, the Maxwell relaxation time is the correlation time of shear-stress autocorrelations.
Calculating the shear viscosity and the high-frequency shear modulus directly from {\it ab initio}
simulations in a wide range of densities, one may calculate the Maxwell relaxation time 
at crossing the MNM transition line. At present it is not clear whether the 
Maxwell relaxation time would show any specific feature at the MNM transition. 
As a consequence of non-monotonous density dependence of $\eta_s$ and monotonous one of $G_{\infty}$, 
the Maxwell relaxation time reveals a drop right in the region of density $\rho=9.25$~g/cm$^{3}$ 
(see figure~\ref{eta}b).

\begin{figure}
	\centerline{\includegraphics[width=0.7\textwidth]{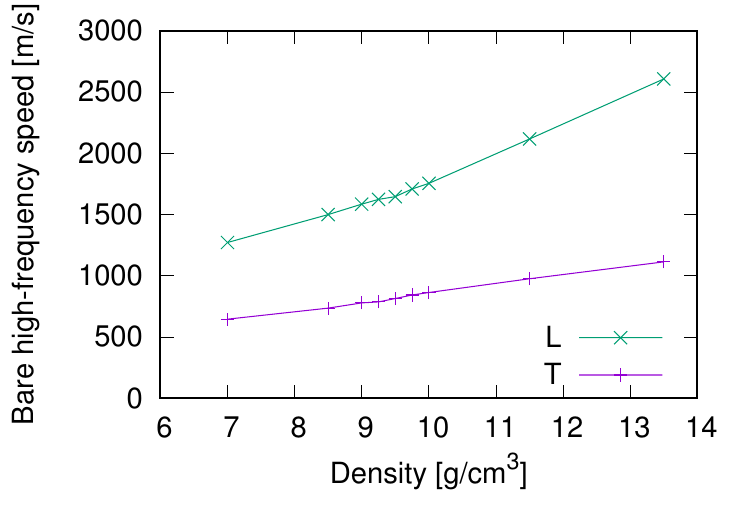}}
	\caption{(Colour online) Density dependence of the high-frequency speeds of 
		non-damped $L$ and $T$ excitations (``bare'' excitations) for liquid Hg at 1750 K. }\label{Figcinf}
\end{figure}

A specific feature of collective dynamics in liquids is the existence of viscoelastic 
transition, when on macroscopic scales (in the long-wavelength region, or, in the hydrodynamic regime) 
only collective processes 
caused by fluctuations of conserved quantities define the dynamic response, while on the spatial 
nano- and shorter scales, the atomistic structure causes a typical elastic response like in solids.
For collective excitations, the viscoelastic transition is observed in the  existence of two characteristic
propagation speeds: adiabatic speed of sound $c_s$ in the long-wavelength limit and high-frequency
speed $c_{\infty}$. The latter can be estimated from simulations using the relation
(\ref{cinf}). For transverse dynamics, the high-frequency speed undamped of transverse excitations
can be calculated as 
$$
c^T_{\infty}=\sqrt{\frac{G_{\infty}}{\rho}}~.
$$
The density dependence of the high-frequency longitudinal and transverse speed of 
``bare'' (undamped) collective excitations in the 
studied density range of fluid Hg is shown in figure~\ref{Figcinf}. One can see that 
both $L$ and $T$ dependencies contain a kind of a kink as a function of the density, corresponding 
to the emergence of the gap in EDOS.

\begin{figure}
		\centerline{\includegraphics[width=0.7\textwidth]{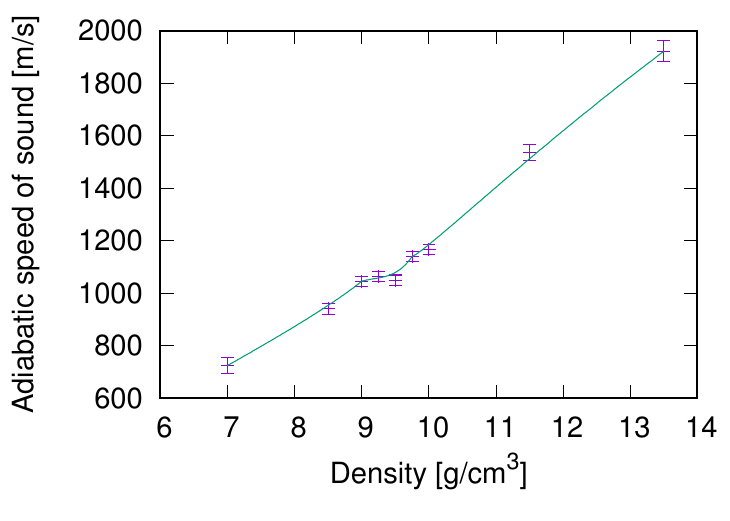}}
\caption{(Colour online) Density dependence of the adiabatic speed of sound for liquid Hg at 1750 K, 
calculated from equation~\ref{cs}.}\label{Figcs}
\end{figure}

The estimation of the adiabatic speed of sound from computer simulations is a demanding task.
Usually, one needs to estimate it from heat density fluctuations sampled in computer simulations,
in particular on the estimates of the $k$-dependent ratio 
of specific heat $\gamma(k)$ and the long-wavelength limit of the smooth $k$-dependence of 
$\gamma(k)/S(k)$ \cite{Bry14b}. However, this methodology is extremely problematic to be applied  
in connection with {\it ab initio} simulations, although a special fitting procedure was 
proposed in reference~\cite{Bry13}. Here, we apply a new approach using a relation (\ref{cs}),
which was shown to yield robust results for many simple and binary liquids \cite{Bry23}.
The density 
dependence of $c_s$ for fluid Hg at 1750~K in figure~\ref{Figcs} shows a plateau in the 
region of the metal-nonmetal transition, a feature mentioned before in the  simulation study \cite{Mun98}
and an analysis of ultrasonic experiments \cite{Oka98}.

\begin{figure}
	\centerline{\includegraphics[width=0.5\textwidth]{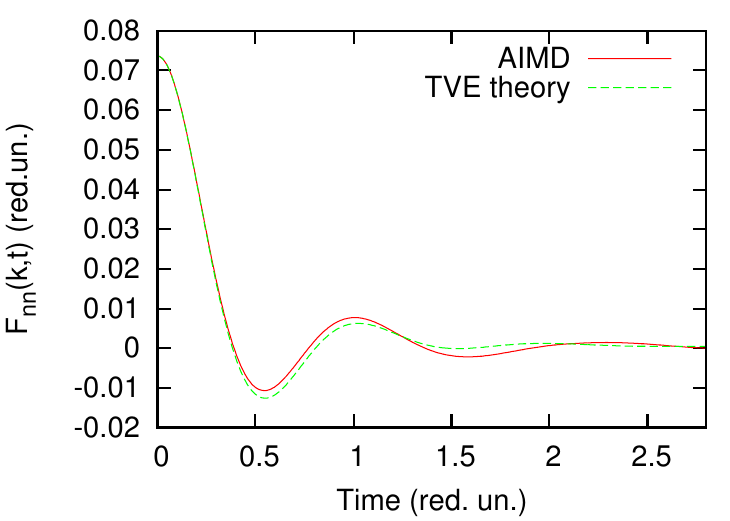}
		\includegraphics[width=0.5\textwidth]{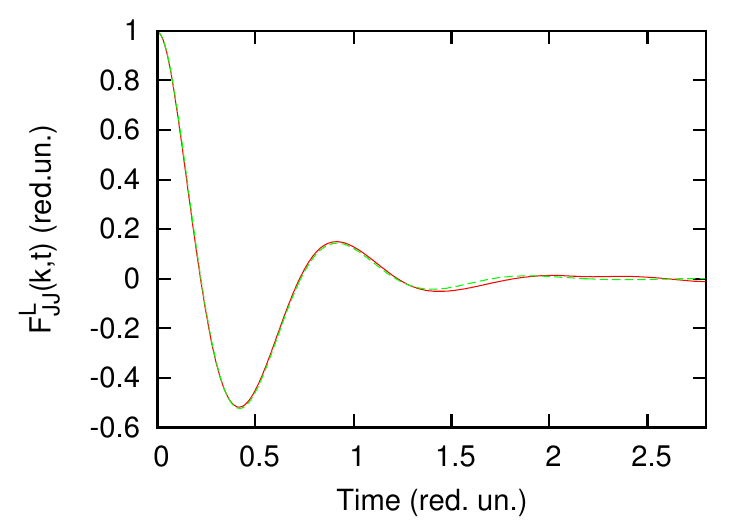}}
\caption{(Colour online) Recovering  the AIMD-derived density-density $F_{nn}(k,t)$ and 
longitudinal current-current $F^L_{JJ}(k,t)$
time correlation functions by the TVE theory for the density 9.5 g/cm$^3$ at wave number 
$k=0.464$~\AA$^{-1}$.}\label{fkt}
\end{figure}

An interesting issue is the behavior of dispersion of acoustic excitations in the region
of MNM transition. As mentioned in the Methodology section, we estimated the dispersion by 
two approaches: purely numerical one via peak positions of the current spectral function
$C^L(k,\omega)$ and a theoretical one via the imaginary part of sound eigenmodes of the 
generalized hydrodynamic matrix ${\bf T(k)}$ at the wave numbers $k$ sampled in AIMD.
A good quality of theoretical description of collective dynamics in liquid Hg can be verified
in figure~\ref{fkt}, where the AIMD-derived density-density and current-current 
time correlation functions at the density 9.5 g/cm$^3$ and wave number 
$k=0.464$~\AA$^{-1}$ are well recovered by the theoretical curves which are separable 
sums of contributions from dynamic eigenmodes of the generalized hydrodynamic matrix 
${\bf T(k)}$ \cite{deS88,Mry95}. 

\begin{figure}[h]
	\centerline{\includegraphics[width=0.48\textwidth]{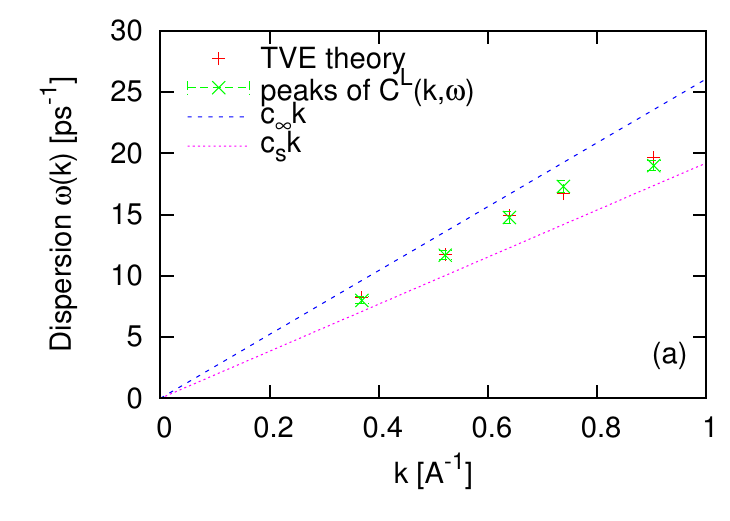}
		\includegraphics[width=0.48\textwidth]{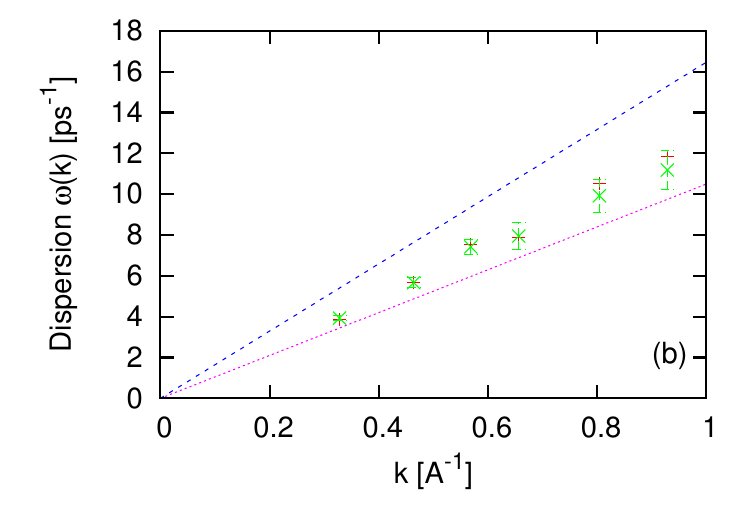}}
\caption{(Colour online) Dispersion of collective excitations in the first pseudo-Brillouin zone for 
two densities of fluid Hg: 13.5 g/cm$^3$ (a) and 9.5 g/cm$^3$ (b). Theoretical 
dispersion via eigenvalues of the thermo-viscoelastic (TVE) model (symbols plus) 
is compared with purely 
numerical estimates via the peak positions of longitudinal current spectral function
$C^L(k,\omega)$ (symbols cross with error bars). The linear dispersion laws with 
high-frequency $c_{\infty}$ and adiabatic $c_s$ speeds of sound are shown by lines.}\label{disp}
\end{figure}

The dispersion of longitudinal collective excitations at two densities, 13.5 g/cm$^3$ and 9.5 g/cm$^3$ 
of fluid Hg, is shown in figure~\ref{disp}. One can see that in the long-wavelength region, the estimated dispersion [plus symbols --- imaginary part of complex eigenvalues of the generalized hydrodynamic matrix and cross symbols with error bars --- peaks of the longitudinal current spectral function $C^L(k,\omega)$] practically matches the linear dispersion law $\omega_{\rm hyd}(k)=c_sk$ with the adiabatic speed of sound estimated from expression (\ref{cs}). Another straight line in figure~\ref{disp} corresponds to a linear dispersion with propagation speed being the high-frequency speed $c_{\infty}$. The theoretical prediction via the dynamic eigenvalues is in very good agreement with the purely numerical estimation of the dispersion of collective excitations via $C^L(k,\omega)$, which was obtained from numerical Fourier transformation of the longitudinal current-current time correlation function and therefore contained the error bars shown in figure~\ref{disp}. For all the studied densities of fluid Hg, we observed a positive deviation of the dispersion of collective excitations from the linear hydrodynamic dispersion law. It is seen in figure~\ref{disp}b that the PSD close to the MNM transition region is $\sim 20$\%. Extended AIMD simulations are needed to reduce the error bars of the apparent dispersion of collective excitations in order to observe the correct behavior of the PSD as a function of density, especially in the region of MNM transition.  

\section{Conclusion}

We performed {\it ab initio} simulations of the fluid Hg at $T=1750$~K covering the 
density range 7.0--13.5~g/cm$^3$. Our simulations and theoretical analysis of collective dynamics 
led to the following conclusions:
\begin{enumerate}[label=\roman*.]
	\item We located the metal-semimetal transition in fluid Hg to occur at density $\rho=9.25$~g/cm$^3$
	as at the density when in the electronic density of states there emerged a gap;
	\item From the behavior of total charge fluctuations, we observed a change in the screening properties 
	of ions by electron density in the region of densities 9.0--9.75 g/cm$^3$. At higher densities,  
	almost perfect metallic screening takes place, while at  densities lower than 9.0 g/cm$^3$, the fluid Hg can be treated as a collection of neutral atoms;
	\item We observed a small plateau in the density dependence of the adiabatic speed of sound, which 
	was located at densities 9.0--9.5 g/cm$^3$, while in the high-frequency speed of sound the plateau 
	was absent, although a possible kink at $\rho=9.25$ g/cm$^3$ can be observed;
	\item The Maxwell relaxation time shows a non-monotonous behavior in the region $\rho=9.0$--9.75 g/cm$^3$, 
	although its changes vs. density are very small;
	\item For all the studied densities of Hg fluid, we observed a positive deviation of the dispersion 
	of collective excitations from the hydrodynamic dispersion law.  
\end{enumerate}

\section*{Acknowledgements}
TB was supported by the National Research Foundation of 
Ukraine Project No. 2023.05/0019. The calculations have been performed using the \emph{ab initio} total-energy and molecular dynamics program VASP (Vienna ab-initio simulation program)
developed at the Institute f\"ur Materialphysik of the Universit\"at Wien
\cite{Kre93,Kre93_a,Kre96,Kre96b}.

{\small \topsep 0.6ex

}

\newpage

\label{ua-part}

\ukrainianpart

\title{Пружні властивості флюїду ртуті в області переходу метал-неметал: Дослідження 
	методом {\it ab initio} моделювання}
\author{Т. Брик\refaddr{label1,label2}, О. Бакай\refaddr{label3}, 
	    А. П. Сейтсонен\refaddr{label4}}
\addresses{
   \addr{label1} Інститут фізики конденсованих систем НАН України, 79011 Львів, Україна,
    \addr{label2} Інститут прикладної математики та фундаментальних наук, Національний університет ``Львівська Політехніка'', 79013 Львів, Україна,
    \addr{label3} Інститут теоретичної фізики ім. Ахієзера, ННЦ ``Харківський фізико-технічний інститут'' НАН України, 61108 Харків, Україна,
    \addr{label4} Хімічний факультет, Вища нормальна школа, 24 вул. Льомон, 75005 Париж, Франція,
}

\makeukrtitle

\begin{abstract}
	\tolerance=3000%
	Ми повідомляємо про дослідження методом {\it ab initio} молекулярної динаміки флюїду ртуті при температурі 1750~К в області густин 7--13.5~г/см$^3$. Вздовж цієї ізотермічної лінії ми виконали аналіз флуктуацій повної зарядової густини, який засвідчив про екранування типу нейтральних атомів у флюїді Hg при густинах менших за 9.25~г/см$^3$, що практично співпадає з появою щілини в електронній густині станів. Високочастотні об'ємний та зсувний модулі, високочастотна та адіабатична швидкості звуку, зсувна в'язкість, Макс\-ве\-лiв\-ський час релаксації та дисперсія колективних збуджень аналізуються як функція густини вздовж ізотермічної лінії.
	\keywords пружні властивості, рідини, поширення звуку, ab initio молекулярна динаміка
	
\end{abstract}

\end{document}